\documentclass[aps,pra,twocolumn,superscriptaddress,showpacs]{revtex4-1}
\usepackage{times,amsmath,amssymb,amsfonts,graphicx,color}

\newcommand{\ket}[1]{\vert #1 \rangle} %

\newcommand{\ketbra}[2]{\vert #1 \rangle \! \langle #2 \vert}
\newcommand\nblp{N_{\rm\scriptstyle BLP}}
\newcommand\nbcm{N_{\rm \scriptstyle BCM}}
\begin{document}
\title{Non-Markovianity of colored noisy channels}
\author{Claudia Benedetti}
\affiliation{Dipartimento di Fisica, Universit\`a degli 
Studi di Milano, I-20133 Milano, Italy} %
\author{Matteo G. A. Paris}
\affiliation{Dipartimento di Fisica, Universit\`a 
degli Studi di Milano, I-20133 Milano, Italy} %
\author{Sabrina Maniscalco}
\affiliation{Scottish Center for Quantum Physics Alliance, Engineering
and Physical Sciences, Department of Physics, Heriot-Watt University,
Edinburgh EH14 4AS, Scotland, United Kingdom }
\date{\today}
\begin{abstract}
We address the non-Markovian character of quantum maps describing the
interaction of a qubit with a random classical field. In particular, we
evaluate trace- and capacity-based non-Markovianity measures for two
relevant classes of environments showing non-Gaussian fluctuations,
described respectively by random telegraph noise and colored noise
with spectra of the the form $1/f^\alpha$. We analyze the dynamics of
both the trace distance and the quantum capacity, and show that the  
behavior of non-Markovianity based on both measures is qualitatively 
similar.  Our results show that environments with a spectrum that contains 
a relevant low-frequency contribution are generally non-Markovian.  We 
also find that the non-Markovianity of colored environments decreases when 
the number of fluctuators realizing the environment increases. 
\end{abstract} 
\maketitle
\section{Introduction}
The unavoidable interaction of a quantum system with its environment
usually destroys its  coherence and quantumness \cite{joos,zurek03}. The
fragile quantum information encoded in an open quantum system is lost
due to the presence of the environment that continuously monitors the
system.  Nevertheless, in some cases the lost information can be partly
restored due to non-negligible correlations between system and
environment. We refer to the systems in which such re-coherence
phenomena occur as non-Markovian open quantum systems.  The dynamics of
open quantum systems has been often described using the Born-Markov
approximation leading to a master equation of the Lindblad form
\cite{bp07}. This approximation, however, neglecting system-bath induced
memory effects,  does not lead to a correct description of the dynamics
of many relevant systems in quantum optics and solid state physics, and
cannot be used in certain quantum information processing scenario
\cite{haikka2011,lorenzo11,liu11,rob11,cos12,lof13}. In addition, in the
spirit of reservoir engineering, one can induce non-Markovianity to
improve quantum protocols such as quantum metrology and quantum key
distribution \cite{huelga,chin,vasile11,lainetelep,bylicka13}.
\par
The concept of non-Markovianity is not uniquely defined in the
literature. Several measures  have been proposed in recent years
\cite{jens,blp,rhp,lu10,luo12,bylicka13,plastina} and, in general, these
measures do not coincide in detecting non-Markovianity.  In this paper
we focus on two measures of non-Markovianity: the Breuer-Laine-Piilo
(BLP) measure \cite{blp} based on state distinguishability, and the
Bylicka-Chru\'sci\'nski-Maniscalco (BCM) measure \cite{bylicka13}
based on entanglement assisted and/or quantum capacities. In the first
case the characteristic trait of non-Markovianity is a backflow of
information, i.e., a partial increase in state distinguishability, while
in the second case memory effects are identified with a regrowth of
channel capacities.  
\par 
It is known that for single qubit dephasing channels, as those
considered in the following, the Markovian or non-Markovian character of
the dynamical map coincides for all measures. Therefore, it is
sufficient to study one of them. In this paper we focus on both the BLP
measure and the BCM measure as we are interested not only in
understanding information back flow but also in investigating under
which conditions qubit channels subject to random classical noise 
{may be exploited for} reliably transmitting quantum and classical information.
Moreover, the BCM measure provides us with a rigorous  information
theoretical description of memory effects by linking the amount of
information on the system to the amount of information on the
environment, and therefore allowing us to properly define the concept of
information flow.  
\par 
We focus on non-Markovianity arising in classical
environments exhibiting non-Gaussian fluctuations, i.e. described by
random non-Gaussian fields. In particular, we address the influence of
this class of environments on the dynamics of a qubit.  As a matter of
fact, little attention has been paid to non-Markovianity in classical
environments, most of the existing studies are devoted to 
time-independent random fields or to
Gaussian dynamic noise \cite{mannone,compagno12,strunz10,strunz11}.  On
the other hand, stochastic processes characterized by non-Gaussian
fluctuations are very common in nature and have received large attention
\cite{galperin06,faoro04,bergli06,paladino07}. A stochastic process is
non-Gaussian if it cannot be fully characterized by the mean and
variance. As a consequence, the mere knowledge of the spectrum is not
sufficient to describe the process, and the very structure of environment 
plays a role in determining its influence on the coherence properties
of quantum systems \cite{benedetti13}.
\par
In this paper we focus on two relevant classes of non-Gaussian 
noise: the random telegraph noise (RTN) with a Lorentzian spectrum 
and the family of low-frequency noise with $1/f^{\alpha}$
spectrum. The RTN is generated from a bistable
fluctuator flipping between two values with a 
switching rate $\xi$. RTN allows one to model
environmental noise appearing in many semiconducting and superconducting
nanodevices \cite{eroms2006,parman91,bergli09,bergli12,benedetti12,
paladino12}. Noises with $1/f^{\alpha}$
spectra are found when the environment can be
described as a collection of  $N_f$ random bistable
fluctuators, with $N_f\geq1$. It affects solid-state
devices, superconducting qubits, and magnetic systems \cite{weissman,
lud84,kogan84,koch07, kaku07,paladino02}.
The dynamical map of a qubit interacting with these kind of environments
describes pure dephasing. In this case the channel is degradable and the
entanglement-assisted capacity coincides with the quantum channel
capacity, hence we will consider only the latter one in the rest of the
paper. Moreover, a simple analytical expressions for the dynamics of
both the trace distance and the quantum capacity exists, allowing us to
analyze in detail the non-Markovian dynamics as a function of the noise
parameters.
Our results show that the two measures display a qualitatively similar
behavior, leading to a consistent assessment of non-Markovianity. We
also find that environments with a spectrum dominated by low-frequency
contribution are generally non-Markovian and that the non-Markovianity of
colored environments is progressively destroyed when the number of
fluctuators realizing the environment increases. 
{These results confirm that non-Markovianity may represent a 
resource for quantum information processing. Indeed, we found 
that, in dephasing channels, back flow of information corresponds to 
revivals of quantum correlations \cite{benedetti13}.}
Finally, {upon assuming
that the channel is reset after each use}, we discuss 
how reliable transmission of information through a noisy quantum channel 
may be achieved, even in presence of classical non-Gaussian noise, 
for properly engineered structured environments.
\par
In principle, one could quantify non-Markovianity of
classical random fields using the definition of non-Markovianity for
classical stochastic processes, i.e. in terms of the Kolmogorov
hierarchy of the $n$-point joint probability distributions. 
In turn, it has been proved \cite{Vac11,bas12}
that there are clear differences between the classical and the quantum notion of
non-Markovianity. In this paper, since we are interested in the {\em
effects} of non-Markovianity on the quantum information carrier rather than in
addressing the fundamental properties of the environment, we have
explicitly chosen to assess non-Markovianity of classical noise in terms
of quantum measures of non-Markovianity for the quantum channels induced 
on qubit systems. This approach allows us to assess classical
environments in terms of their effects on quantum systems and, possibly, 
engineer their structure.
\par
This paper is organized as follows. In Sec. \ref{sec1} we introduce 
the BLP and BCM measures of non-Markovianity. In Sec. 
\ref{sec2} we describe in some details the physical model employed
throughout the paper. In Sec. \ref{sec3} we show our 
results on the dynamics of the non-Markovianity measures. 
Section \ref{sec4} closes the paper with concluding remarks.
\section{Non-Markovianity measures}\label{sec1}
In this section we review two measures of non-Markovianity:
the BLP measure \cite{blp}, based on the trace distance, and the 
BCM measure, based on the quantum capacity \cite{bylicka13}.
\subsection{BLP measure}
The underlying idea behind the BLP measure is that Markovian processes
tend to reduce the distinguishability between any two states of the open system,
while non-Markovian processes are characterized by a partial re-growth 
of distinguishability on at least a subset of states. 
The loss of distinguishability is interpreted
as an irreversible loss of information on the system while restored distinguishability reflects 
a partial, and often temporary, increase of information about the system.
Since the trace distance is related to the
probability of successfully distinguishing two quantum states $\rho_1$ and 
$\rho_2$, it
seems natural to use this quantity to describe memory effects and non-Markovianity. 
The trace distance is defined as:
\begin{equation}
 {D}(\rho_1,\rho_2)=\frac{1}{2}\text{Tr}|\rho_1-\rho_2|,\label{trdist}
\end{equation}
where $|A|=\sqrt{A^{\dagger}A}$. The trace distance defines 
a metric on the space of density matrices  and $0<{D}<1$.
Any completely positive and trace preserving map $\Phi(t)$
is a contraction for this metric:
\begin{equation}
{D}(\Phi(t)\rho_1,\Phi(t)\rho_2)<{D}(\rho_1,\rho_2).
\end{equation}
The flux of information is then defined as:
\begin{equation}
 \sigma(t,\rho_{1,2}(0))=\frac{d}{dt}{D}(\rho_1,\rho_2).\label{flux}
\end{equation}
where $\rho_{1,2}(0)$ denotes the density matrix
of the initial states.
A loss in distinguishability is linked to a negative information flux 
$\sigma(t,\rho_{1,2}(0))<0$, while positive flux describes
an increasing distinguishability between two quantum states.
The BLP measure of non-Markovianity quantifies the total amount of 
information backflow into the system:
\begin{equation}
 \nblp=\max_{\rho_{1,2}(0)}\int_{\sigma>0}ds\,\sigma(s,\rho_{1,2}(0))\label{blpm}
\end{equation}
where the maximization is over all pairs of initial states and the
integration is over all time intervals where the flux is positive.
Whenever $\nblp>0$ the dynamics is non-Markovian.
It is possible to show that all divisible maps are Markovian according
to this measure, but the converse is not true in general \cite{LPB2}.
Generally, calculating $\nblp$ is a difficult task, because of 
the maximization procedure involved. Nevertheless, maximizing pairs have been 
found for certain classes of noisy channels, such as the dephasing channel we
are going to deal with. As we will review in Sec. III, indeed, the dephasing dynamics of the qubit is described by the following map
\begin{equation}
\rho(t)=\frac{1+\Gamma(t)}{2}\rho(0)+\frac{1-\Gamma(t)}{2}\sigma_z\rho(0),
\sigma_z.\label{deph}
\end{equation}
i.e., $\rho_{ij}(t)=\rho_{ij}(0) \Gamma(t)$, with $\Gamma(t)$ the dephasing function.
Notice that in Eq. \eqref{deph} dephasing is induced by 
a classical stochastic process. It follows that the coefficient $\Gamma(t)$ is a real quantity that 
can assume positive and negative values between $\pm1$. For the sake of clarity we notice here the difference between this model and the spin-boson model describing pure dephasing arising from the interaction with a quantum environment \cite{dephasing,haikka13}. In the latter case the dynamics of the coherences is given by $\rho_{ij}(t)=\rho_{ij}(0) \exp [- f(t)]$, with $f(t) \ge 0$, and therefore the dephasing function is always positive.
\par
For a qubit interacting with a classical random field, as in Eq. \eqref{deph}, 
the optimal pair in Eq. \eqref{blpm} that maximizes
the increase of the trace distance has been found \cite{shao}. In the following, 
we use the name {\em optimal trace distance} for the expression
of the trace distance in Eq. \eqref{trdist} computed for the optimal pair.
Thus, the 
study of BLP non-Markovianity can be reduced to the analysis of 
the {optimal trace distance}.
In our case this quantity reads: 
\begin{equation}
 {D}(t)=|\Gamma(t)|,\label{maxtd}
\end{equation}
which is the absolute value of the dephasing function $\Gamma(t)$. Once the expression
of ${D}(t)$ is known it is possible to
write the information flux \eqref{flux} and then numerically compute the non-Markovianity
as the integral of the information flux over the time intervals in which it is positive.
\subsection{BCM measure}
The second measure that we will use, introduced by Bylicka,
Chru\'sci\'nski and Maniscalco,  is based on the channel capacities
\cite{bylicka13}. As mentioned above, we focus here on the quantum
channel capacity as, in our case, it coincides with the entanglement
assisted capacity.  Generally, the quantum capacity bounds the rate at
which quantum information can be reliably transmitted through a noisy
quantum channel $\Phi(t)$.  Dephasing is a degradable channel
and thus the single-use quantum capacity may be written as 
\begin{equation}
 C_Q(\Phi(t))=\sup_{\rho} I_c(\rho,\Phi(t))\label{qc1}
\end{equation}
where $I_c$ represents the coherent information:
\begin{equation}
 I_c(\rho,\Phi(t))=S(\Phi(t)\rho)-S(\rho,\Phi(t)),
\end{equation}
where $S(\rho) = - \hbox{Tr} \left(\rho \ln \rho \right)$
is the von-Neumann entropy of the state $\rho$ and $S(\rho,\Phi(t))$
the entropy exchange \cite{shumacher}.  The superior should be
taken over all the possible states of the information carrier. 
The latter quantity measures the
change in the entropy of the environment. It is worth stressing that,
contrarily to trace-distance, the coherent information describes how the
entropy of both the system and the environment change.  Hence this
quantity is more apt to capture the concept of information flow between
system and environment.  More explicitly, coherent
information is, in fact, quantifying the flow of information {\rm
between} the system and the environment, while the trace distance is 
quantifying only the {\em loss} of information that we have {\em on 
the system} without any indication on whether this information is 
acquired by another agent (and may come back). 
The fact that the two
measures agree confirms, at least for dephasing channels, that the trace 
distance may be considered a measure of
the flow of information from the system to the environment.
\par
{Strictly speaking, Eq. (\ref{qc1}) describes the quantum capacity 
only for memoryless degradable channels \cite{arr07,vir07,car12}
and thus it seems unsuitable to address quantum channels arising from 
the interaction with a classical enivironment exhibiting long 
lasting time correlations. However, we are interested in the effects
of memory {\em during the propagation} of the information carriers, 
rather than memory effects among subsequent uses of the channel, and 
thus the expression in (\ref{qc1}) fits nicely for our purposes.
Of course, memory effects among uses should be eventually addressed 
in view of realistic implementations. Our results should be
considered as a first step toward a complete analysis of this class of
channels, including both types of memory effects \cite{sjpb}.}
\par
As a consequence of the quantum data-processing inequality, the quantum
capacity of divisible channels is a monotonically decreasing function of
time. The BCM measure is therefore based on the non-monotonic behavior
of $C_Q(\Phi(t))$: \begin{equation}
 \nbcm=\int_{\frac{dC_Q(\Phi(t))}{dt}>0}\frac{d\,C_Q(\Phi(t))}{dt}\,dt.
\label{bcmm}
\end{equation}
Non-Markovianity corresponds to $\nbcm>0$.  Even if this
measure does not explicitly involve a maximization procedure, an
optimization is required in the computation of the quantum capacity
\eqref{qc1}.  In the case of a dephasing channel, however, the
analytical expression for $C_Q(\Phi(t))$ is known \cite{wilde}:
\begin{equation}
 C_Q(t)=1-H_2\left(\frac{1-\Gamma(t)}{2}\right),\label{qcap}
\end{equation}
with $H_2(p)=-p\log p-(1-p)\log(1-p)$ the Shannon binary entropy.  As we
will see in the following, the operational interpretation of the quantum
capacity allows us to use $\nbcm$ to assess if and how
non-Markovianity can be seen as a resource for quantum communication and
information processing.
\par
Both the BLP and the BCM measure detect consistently non-Markovianity in
the case of a dephasing channel.
It is easy to show that in both cases the dynamics is non-Markovian
in the following two regimes:
\begin{equation}
 \begin{array}{c}
  \dfrac{d\Gamma(t)}{dt}<0 \text{ and }\Gamma(t)<0\\
  \\
 \dfrac{d\Gamma(t)}{dt}>0 \text{ and }\Gamma(t)>0
 \end{array}
\end{equation}
where $\Gamma(t)$ is the real coefficient appearing in
Eq. \eqref{deph}
\section{The physical model}\label{sec2}
In order to characterize the non-Markovian features of a noisy channel
one has to probe its influence on an information carrier. Here we  
focus attention on the simplest quantum probe, i.e. a qubit, and infer 
the properties of the channel by analyzing the decoherence process 
induced on the qubit. We model the classical environment as a random 
field described by a stochastic process $c(t)$. In order to model the 
colored noisy channel we do not make the usual assumption of a 
Gaussian process, while we only assume the stationarity property.
\par
Let us consider a generic qubit described by the state vector
\begin{equation}
\ket{\psi_0}=\alpha\ket{0}+\beta\ket{1},\label{state}
\end{equation}
satisfying the condition $|\alpha|^2+|\beta|^2=1$.
We assume that the evolution of the qubit is governed 
by the  Hamiltonian 
\begin{equation}
 H(t)=\epsilon\; \sigma_z +\nu\; c(t) \sigma_z,\label{hamiltonian}
\end{equation}
where $\epsilon$ is the energy splitting between the two levels of the
qubit, $\nu$ is a
coupling constant, and $\sigma_z$ is the Pauli matrix. 
The system-environment interaction of Eq.  (\ref{hamiltonian}) describes
a non-dissipative dephasing channel, and it is suitable to portray
situations where the typical frequencies of the environment are smaller 
than the natural frequency of the qubit.
\par
Different expressions for $c(t)$ may be employed, corresponding to different 
kinds of classical noise. In the next sections, we briefly review the
dynamics of the generic qubit state \eqref{state} in environments 
characterized by RTN and colored noise spectra of the form
$\frac{1}{f^{\alpha}}$. Then, in Sec. \ref{sec3} we will employ 
these results to explicitly evaluate the non-Markovianity of 
the corresponding channels.
\subsection{Random telegraph noise}
Random telegraph noise is very common in nature. It appears in semiconductor,
metal and superconducting devices \cite{eroms2006,parman91,bergli09,bergli12,benedetti12,
paladino12}.
The source of random telegraph noise is a bistable fluctuator, which is a 
quantity which flips between two values with a switching rate, such as
a resistance switching between two discrete values, charges jumping
between two different locations or electrons that flip their spin.
In order to describe a RTN, the quantity $c(t)$ in Eq. \eqref{hamiltonian}
flips randomly between values $\pm1$ with a switching
rate $\xi$. This noise is characterized by an exponential decaying
autocorrelation function $C(t,t_0)$ and a Lorentzian spectrum $S(\omega)$:
\begin{align}
 C(t,t_0)=e^{-2\xi\,|t-t_0|}, \label{autocorrelation} \\
 S(\omega)\propto \frac{4\xi}{4\xi^2+\omega^2}.
\end{align}
Hereafter we use dimensionless quantities. In particular we introduce the dimensionless
time $\tau\equiv\nu t$ and switching rate $\gamma\equiv\xi/\nu$.
The dynamics of the global system 
is described in the interaction picture by the evolution
operator $U(\tau)=e^{-i \varphi(\tau)\sigma_z}$ where 
$$\varphi(\tau)= \int_0^\tau c(s)ds$$ is the RTN phase \cite{benedettiIJ}.
The qubit density matrix is obtained as the average
of the evolved density operator over the process, i.e. over 
the RTN phase:
\begin{equation}
 \rho(\tau,\gamma)=\langle U(\tau)\rho_0 U^{\dagger}(\tau)\rangle_{\varphi(\tau)}\label{mrtn}
\end{equation}
where $\rho_0=\ketbra{\psi_0}{\psi_0}$. The qubit density matrix
thus has the expression:
\begin{equation}
 \rho(\tau,\gamma)=\left(\begin{array}{cc}
                             |\alpha|^2&\alpha\beta^{*}\;G(\tau,\gamma)\\
                             \alpha^{*}\beta \;G(\tau,\gamma)&|\beta|^2
                            \end{array} \right)\label{rhor}
\end{equation}
where the $z^*$ denotes the conjugate of the complex number $z$. 
The coefficient $G(\tau,\gamma)=
\langle e^{2i\varphi(\tau)}\rangle$ corresponds to the function 
$\Gamma(t)$ in Eq. \eqref{deph}, and is given by:
\begin{equation}
 G(\tau,\gamma)=e^{- \gamma \tau }\left(\cosh\delta\tau+
\frac{\gamma \sinh\delta\tau}{\delta}\right), \label{GRTN}
\end{equation}
with $\delta=\sqrt{\gamma^2-4}$.
\subsection{Colored noise}
Power-law frequency noise characterized by $1/f^{\alpha}$ spectrum is an 
ubiquitous noise in nature \cite{weissman}. It can be found in nanoscale 
electronic devices where it manifests as charge fluctuations \cite{lud84,kogan84} and 
in Josephson circuits due to fluctuating background charges and flux
\cite{koch07,kaku07,paladino02}. Typical values of the coefficient $\alpha$
range between 0.5 and 2.
$1/f^{\alpha}$ noise arises
when the environment is described as a collection of bistable 
fluctuators \cite{benedetti13}. Every fluctuator has an unknown
switching  rate, taken from the ensemble $\{\gamma_i,p_{\alpha}(\gamma_i)\}$,
where the probability distribution is:
\begin{align}\label{distrib}
p_{\alpha}(\gamma)=\left\{\begin{array}{ll}
 \frac{1}{\gamma\,\ln(\gamma_2/\gamma_1)}&\alpha=1\\
 & \\
\frac{\alpha-1}{\gamma^{\alpha}}
 \left[\frac{(\gamma_1\gamma_2)^{\alpha-1}}
 {\gamma_2^{\alpha-1}-\gamma_1^{\alpha-1}}\right]&\alpha \neq1
\end{array}
\right.
\end{align}
The coefficient $c(t)$ in Hamiltonian \eqref{hamiltonian}
is a linear superposition of $N_f$ random bistable fluctuators,
$c(t)=\sum_i^{N_f}c_i(t)$, where every $c_i(t)$ describes a stochastic
telegraphic process with a Lorentzian spectrum. 
Following Ref. \cite{benedetti13}, we can write the global
noise phase as the product of RTN phases, $\varphi(\tau)=
\prod_i^{N_f}\varphi_i(\tau)$. The evolution operator
in the interaction picture is written as $U(\tau)=e^{-i\varphi(\tau)\sigma_z}$.
The qubit density matrix is obtained as an average of the global density matrix 
$U(\tau)\rho_0U^{\dagger}(\tau)$ 
over the total noise phase and the switching rates:
\begin{equation}\label{integ}
\rho(\tau)=\int_{\gamma_1}^{\gamma_2}\rho(\tau,\gamma) \;p_{\alpha}(\gamma)\;d\gamma
\end{equation}
where now $\rho(\tau,\gamma)$ has the
same expression as in Eq. \eqref{mrtn} but the average is over the global phase,
$\gamma_1$ and $\gamma_2$ are the smallest and the
biggest switching rate considered.
The qubit density matrix can thus be written as:
\begin{align}
\rho(\tau)=\left(\begin{array}{cc}
                     |\alpha|^2&\alpha\beta^{*}\;\Lambda(\tau,\alpha,N_f)\\
                     \alpha^{*}\beta\;\Lambda(\tau,\alpha,N_f)&|\beta|^2
                    \end{array}\right).\label{rhoc}
\end{align}
where 
\begin{equation}
 \Lambda(\tau,\alpha, N_f)=\left[\int_{
 \gamma_1}^{\gamma^2} G(\tau,\gamma)\; p_{\alpha}(\gamma) \;
d\gamma\right]^{N_f}\,.\label{lambda}
\end{equation}
Also in this case the dephasing factor 
$\Lambda(\tau,\alpha,N_f)$ in Eq. \eqref{rhoc}
corresponds to the coefficient $\Gamma(t)$ in Eq. \eqref{deph}.
\begin{figure}[t]
\centering
\includegraphics[width=0.9\columnwidth]{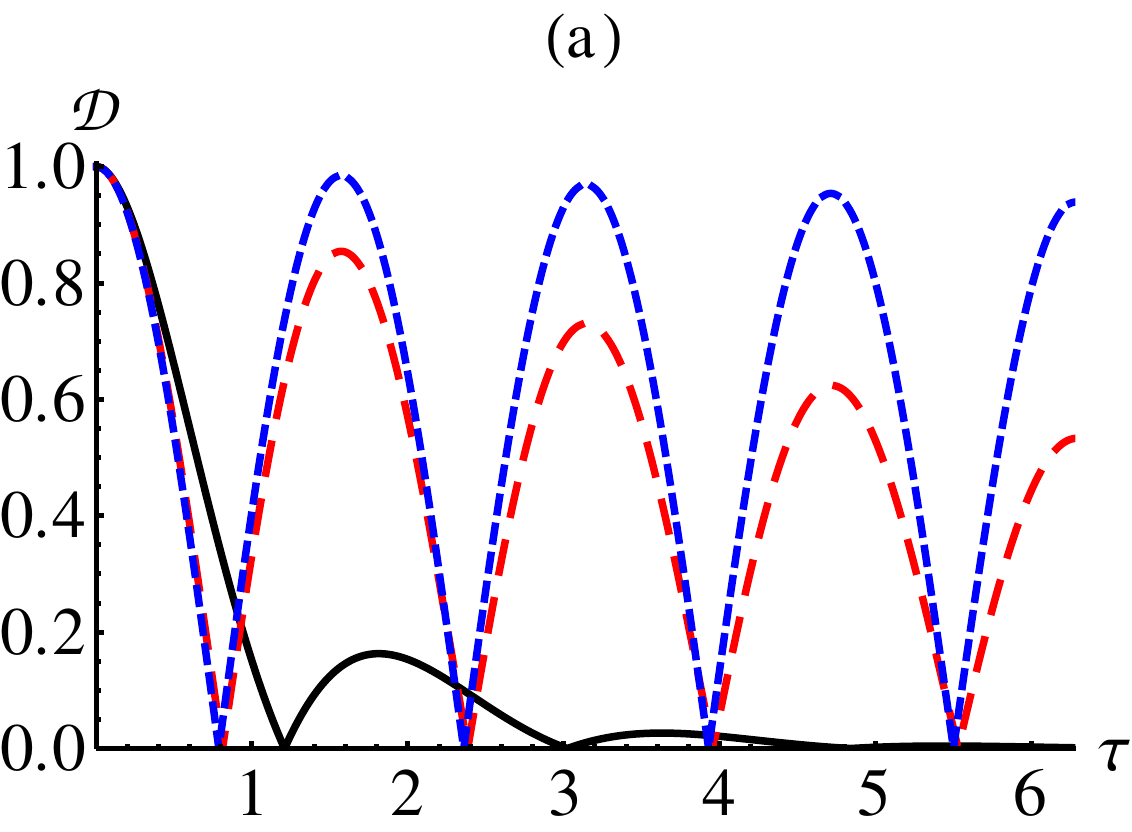}\\
\includegraphics[width=0.93\columnwidth]{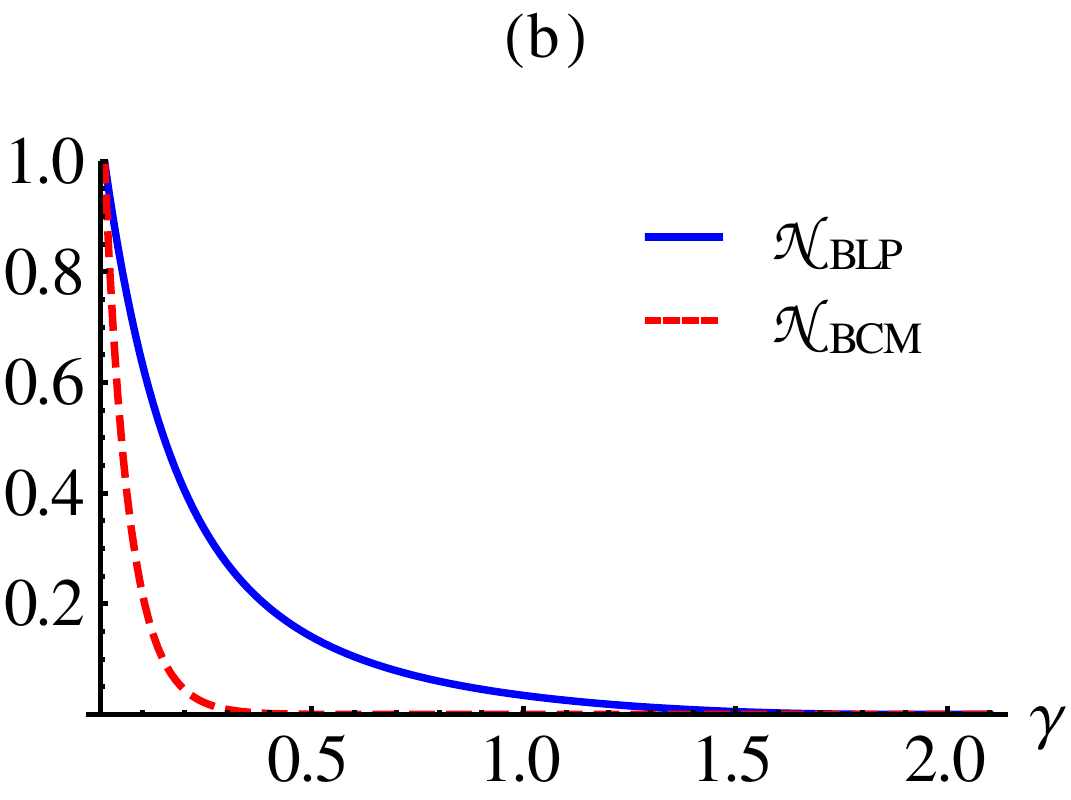}
\caption{(Color online) Non-Markovianity of RTN channels. 
The upper panel shows the trace 
distance as a function of time for three different values 
of the switching rate: $\gamma=1$ (solid black line),
$\gamma=0.1$ (dashed red line) and $\gamma=0.01$ (dotted blue line).
The lower panel is a log-plot of both BLP and BCM non-Markovianity measures 
as a function of $\gamma$. } \label{fig1}
\end{figure}
\section{Non-Markovianity of non-Gaussian noisy channels}
\label{sec3}
The trace distance as well as the quantum capacity
 depends only on the dephasing factor in the density matrices
in Eqs. \eqref{rhor} and \eqref{rhoc}.  Thus their behavior,
even if quantitatively different, is qualitatively the same. 
It immediately follows that also $\nblp$ and 
$\nbcm$ depend only on the dephasing factor and
are characterized by the same behavior.
For any given values of the interaction time,  $\nblp$ 
and $\nbcm$ are functions of noise  parameters describing
the channels.
In the case of RTN, there is a single parameter, i.e. the switching rate
$\gamma$, whereas for colored noise with $1/f^{\alpha}$ spectra the 
tunable parameters are the exponent $\alpha$ and the number of fluctuators.
\par
In this section we study the non-Markovian character of qubits subjected
to RTN and coloured noise. In the spirit of reservoir engineering this
allows one to single out the values of noise parameters minimizing
dephasing and/or leading to to environment-induced re-coherence, i.e.
non-Markovianity.  Similarly, we will see how the  study of the behavior
of the quantum capacity as a function of time reveals the existence of
specific  channel lengths (interaction times) permitting a reliable
transmission of information even in presence of high levels of noise.
Finally, we will conclude our analysis with the characterization of
non-Markovianity of classical environments acting on two-qubit states
and show that the non-Markovianity is quantitatively the same as the
single-qubit case for both measures.
\subsection{Random telegraph noise}
The first step to compute the BLP and BCM non-Markovianity measures is the
evaluation of the optimal trace distance and the quantum capacity, respectively. 
From Eqs. \eqref{maxtd} and \eqref{qcap}, we can write:
\begin{align}
 {D}(\tau,\gamma)&=|G(\tau,\gamma)|,\\
 C_Q(\tau,\gamma)&=1-H_2\left(\frac{1-G(\tau,\gamma)}{2}\right),
\end{align}
with $G(\tau, \gamma)$ given by Eq. \eqref{GRTN}. Two different regimes
naturally arise: for $\gamma<2$ both the trace distance and the quantum
capacity display damped oscillations, i.e. the dynamics is
non-Markovian, whereas for $\gamma\geq 2$ they decay monotonically, i.e.
the dynamics is Markovian. 
In fact, the non-Markovianity measures $\nblp$ and $\nbcm$ correspond
to the integrals, over their range of positivity, of the 
quantities
\begin{align}
\sigma_{\rm \scriptstyle BLP} =& 
\frac{d}{d\tau} D(\tau,\gamma) = 
- \gamma \left|G(\tau,\gamma)\right| \notag \\ 
&+
\hbox{sgn}\left[G(\tau,\gamma)\right] \left(
\gamma \cosh \delta\tau +\delta\sinh \delta\tau
\right) e^{-\gamma\tau} \notag \\
&\qquad\qquad\;\,\, = - 4\, e^{-\gamma\tau}\, \frac{\sinh
\delta\tau}{\delta} \qquad \hbox{if } \gamma\geq 2
\notag \\
\sigma_{\rm \scriptstyle BCM}  = &
\frac{d}{d\tau} C_Q(\tau,\gamma) = -\frac{4}{\ln 2} 
\,\frac{\sinh \delta\tau}{\delta}
\,\hbox{arctanh}\, G(\tau,\gamma)
\,,\label{sigmas}
\end{align}
respectively, 
where $G(\tau,\gamma)$ is given in Eq.(\ref{GRTN}). As it is apparent
from their expressions, and from the fact that $G(\tau,\gamma) \geq 0$
for $\gamma\geq 2$, both the $\sigma$'s are negative definite for
$\gamma\geq 2$, such that both the measures $\nblp$ and $\nbcm$ vanish
for $\gamma\geq 2$. On the other hand, both the $\sigma$'s show an
oscillatory behavior as a function of time, which includes positive 
values, for any values of $\gamma <2$.
\par
Using Eq. (\ref{sigmas}) one finds the extrema of the functions $D(\tau,
\gamma)$ and $C_Q(\gamma,\tau)$ and thus the regions where they are
increasing function of time. Maxima are located at
$\tau_k=k\pi/\sqrt{4-\gamma^2}$ and minima (where the two functions
vanish) at $\tau_k-\tau^*$, with
$\tau^*=(4-\gamma^2)^{-\frac12}\hbox{arctanh}
[\gamma(4-\gamma^2)^{\frac12}]$. We thus obtain
\begin{align}
\nblp =& \sum_{k=1}^\infty D(\gamma,\tau_k) 
     =\left[\exp\left(\frac{\pi\gamma}{\sqrt{4-\gamma^2}}\right)-1\right]^{-1}
     \\
\nbcm =& \sum_{k=1}^\infty C_Q(\gamma,\tau_k)  
\end{align}
In the upper panel of Fig. \ref{fig1} we show the trace distance dynamics in the 
non-Markovian regime $\gamma<2$ for three specific values of 
$\gamma$. We notice that the smaller is $\gamma$, the higher 
are the revivals of the trace distance, and thus the more enhanced is the 
non-Markovian character of the dynamics. A similar behavior 
occurs for the quantum capacity. Indeed, one can see from 
the lower panel of Fig. 
\ref{fig1} that both $\nblp$ and $\nbcm$ 
increase for decreasing values of $\gamma$. 
From a physical point of view this reflects the fact that small values 
of $\gamma$ correspond to non-negligible and long-living environmental 
correlations, as described by the autocorrelation function of 
Eq. \eqref{autocorrelation}, 
and therefore to more pronounced memory effects. 
The lower panel of Fig. \ref{fig1} also shows 
that $\nbcm$ decays faster than $\nblp$ as a function of $\gamma$.
On the other hand, as mentioned above, the threshold between the
Markovian and non-Markovian regime is the same for both measures and
corresponds to $\gamma=2$, for which both measures vanish. 
More precisely, for $\gamma\geq 2$ both the measures are identically 
zero since the time derivatives of both the trace distance and the
quantum capacity are negative definite, meaning that
information permanently leaks away from the system. 
%
%
\subsection{Colored $1/f^{\alpha}$ noise}
For the sake of conciseness we focus initially on the behavior of trace distance, as the dynamics of the quantum capacity is qualitatively similar. 
For colored noise with spectrum $1/f^{\alpha}$, the optimal trace distance
is:
\begin{equation}
{D}(\tau,\alpha,N_f)=|\Lambda(\tau,\alpha,N_f)|.\label{Dalph}
\end{equation}
This quantity cannot be evaluated analytically since the integral in 
Eq. \eqref{lambda} is not analytically solvable. We computed it numerically 
upon assuming that the range of  integration in Eq. (\ref{lambda}) includes 
rates belonging to the interval  $[\gamma_1,\gamma_2]=[10^{-4},10^4]$. 
\par
The optimal trace distance for a generic number
of fluctuators may be written in terms of the same quantity 
for a single fluctuator as follows 
\begin{equation}
{D}(\tau,\alpha,N_f)={D}(\tau,\alpha,1)^{N_f} \label{DNF}.
 \end{equation}
We thus first analyze the non-Markovianity of a colored 
environment generated by a single random fluctuator.
\begin{figure}[h!]
\centering
\includegraphics*[width=0.8\columnwidth]{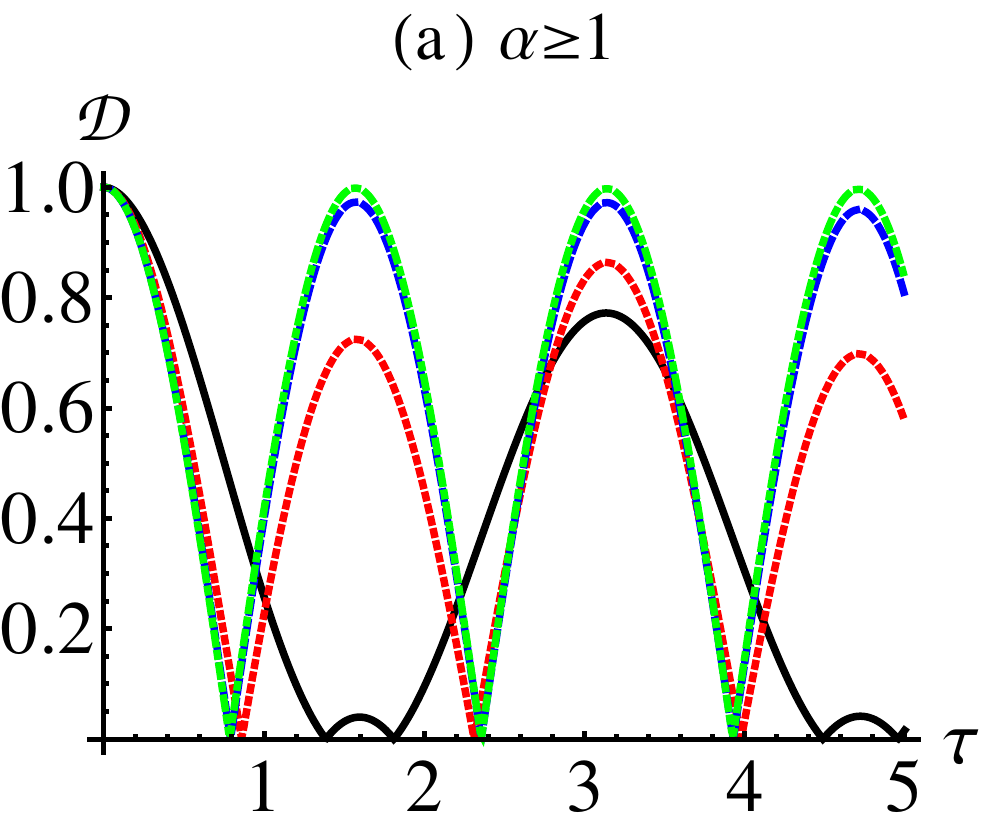}\\
\includegraphics*[width=0.8\columnwidth]{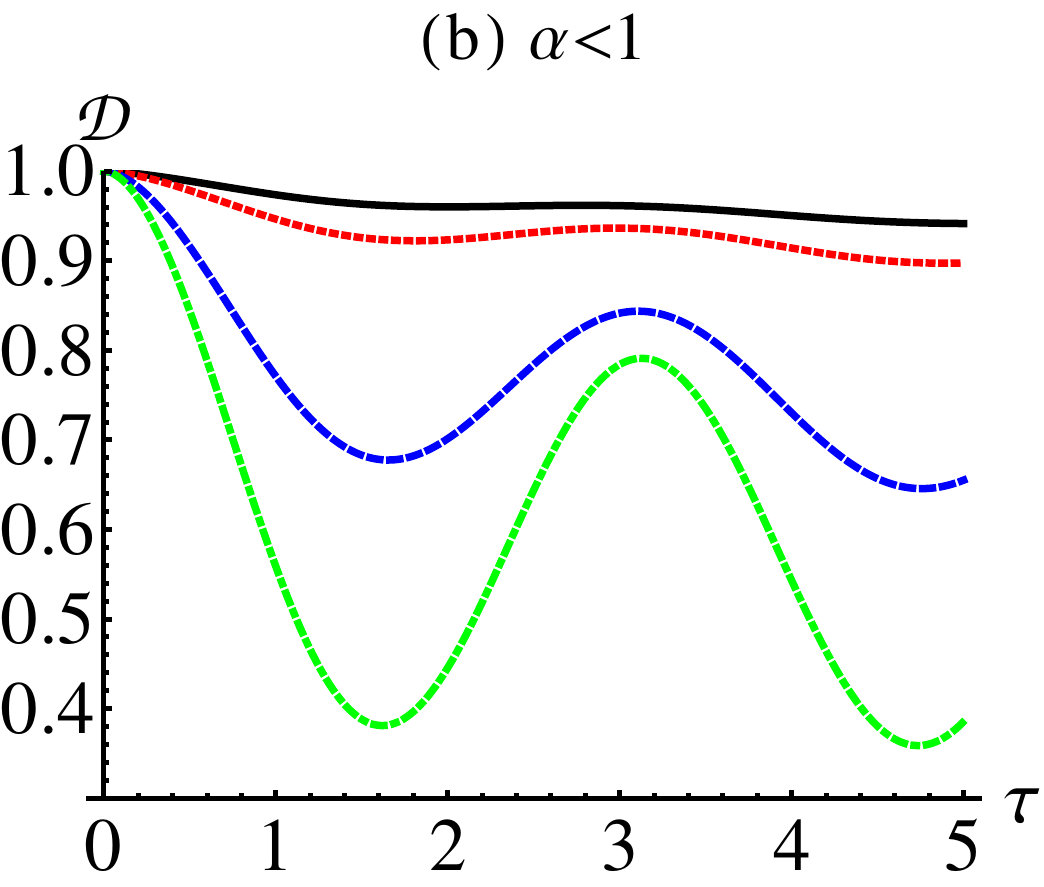}
\caption{(Color online) Non-Markovianity of colored channels. The 
upper panel shows the trace distance for a qubit subject to
$1/f^{\alpha}$ noise generated by a single random 
fluctuator for different values of $\alpha\geq1$: $\alpha=1$ (solid black line),
$\alpha=1.3$ (dotted red line), $\alpha=1.5$ (dashed blue line) and 
$\alpha=2$ (dotdashed green line). The lower panel shows the same
quantity for values of $\alpha<1$: $\alpha=0.5$ (solid black line),
$\alpha=0.6$ (dotted red line), $\alpha=0.8$ (dashed blue line) and 
$\alpha=0.9$ (dotdashed green line). }\label{fig2}
\end{figure}
For a fixed value of $\alpha$ and $N_f=1$ the trace distance is always a
non-monotonic function  of time, as illustrated in Fig. \ref{fig2}.
Therefore, contrarily to the case of RTN, the dynamics is always
non-Markovian for a single fluctuator. However, one can still identify
two regimes depending on the value of $\alpha$.  
\par
\begin{figure}[h!]
\centering
\includegraphics*[width=0.8\columnwidth]{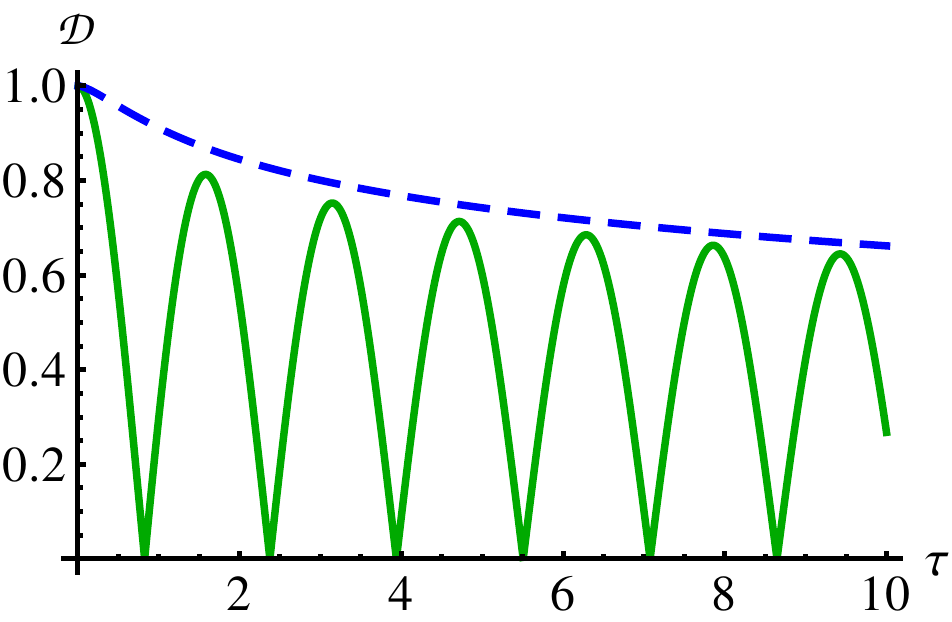}
\caption{(Color online) Non-Markovianity of colored channels. The plot
shows the trace distance as a function of time for $\alpha=1$. The two
curves refer two different ranges of integration in Eq. (\ref{lambda}): 
$[10^{-4},2]$
(green solid line) and $[2,10^{4}]$(blue dashed line) }\label{fig3}
\end{figure}

 
For, $\alpha\geq1$ the optimal trace distance is characterized by
pronounced oscillations in time between zero and a maximum value, which
depends upon the value of $\alpha$.  The larger is $\alpha$, the larger
are the local maxima. For $\alpha=1$ higher and lower maxima alternate
periodically at times $\tau=\pi/2$. As $\alpha$ increases, the height of
the alternating peaks increases until it becomes uniform, as shown in
Fig. \ref{fig2} (a).  For $\alpha<1$, ${D}$ is still non-monotonic, but
the oscillations are less noticeable and the optimal trace distance
never vanishes.  This case is illustrated in  Fig. \ref{fig2} (b).
Generally, as $\alpha$ decreases, the amplitude of the oscillations in
the trace distance decreases, both in the $\alpha<1$ and in the
$\alpha\geq1$ region of parameter space.  Non-Markovianity is thus
stronger for systems interacting with environments with a dominant  low
frequency component in the frequency spectrum.  By changing the range of
integration in Eq. \eqref{lambda}, we can analyze the contribution of
small and large switching rates on non-Markovianity. In particular, in
Fig. \ref{fig3} we compare the behavior of the trace distance for two
mutually exclusive ranges of integration, namely for $\gamma \in
[10^{-4}, 2]$ and $\gamma \in [2,10^{4}]$.  In the first case we
integrate only over small values of the switching rates, and the trace
distance exhibits revivals, revealing the presence of information back
flow.  In the second case the integration is performed over big values
of $\gamma$ and, as a result, ${D}(t,\alpha,1)$ decays monotonically.
This behavior is consistent with the results obtained for the RTN
channel: memory effects are dominant for low switching rates, i.e.
longer correlation times. Thus non-Markovianity is a distinctive trait
of low-frequency noise spectrum.  
\par
\begin{figure}[h!]
\centering
\includegraphics*[width=0.9\columnwidth]{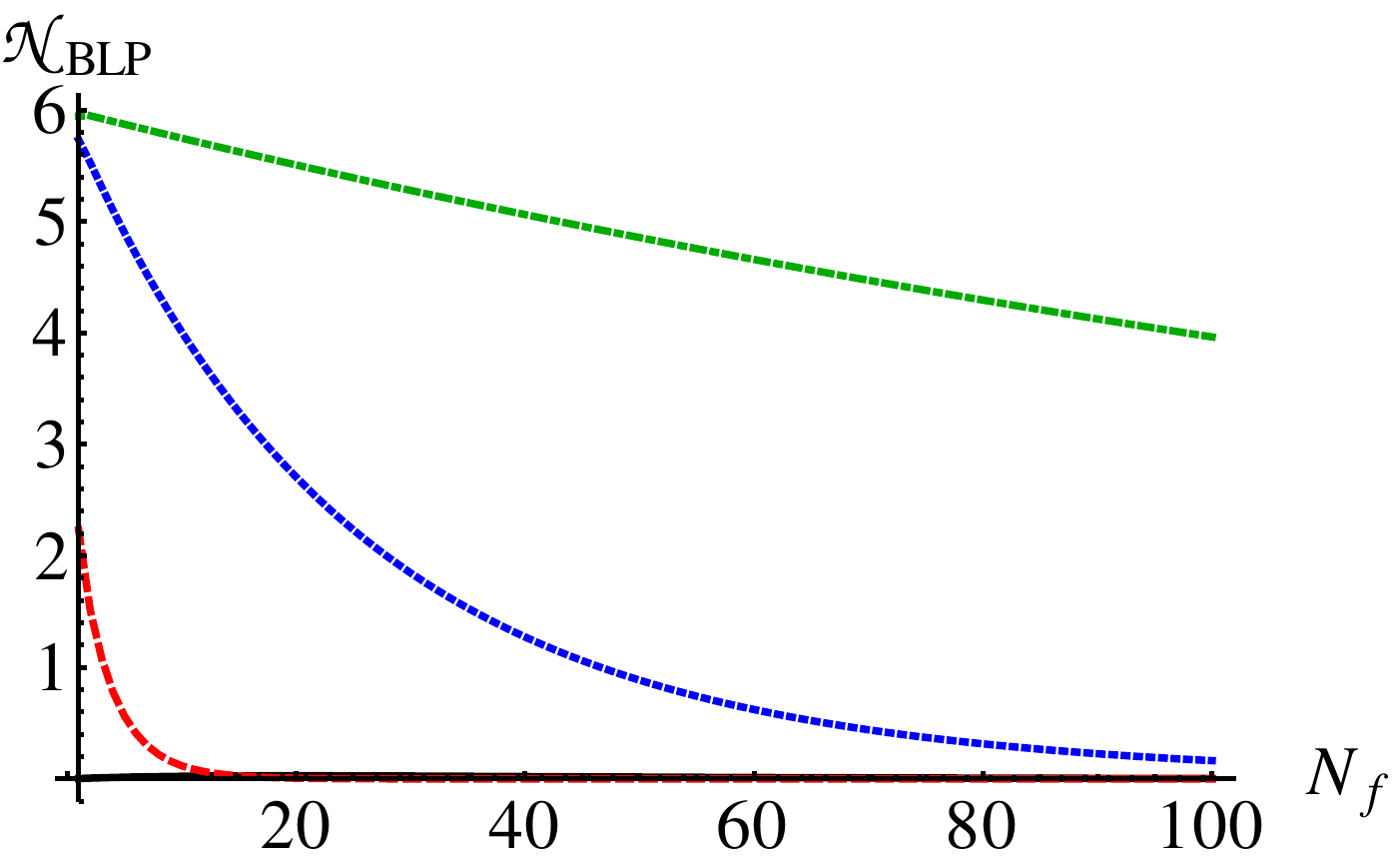}
\caption{(Color online) Non-Markovianity of colored channels. The 
plot shows the non-Markovianity measures as a function of the
number of fluctuators. The plot shows $\nblp$ (solid lines) and $\nbcm$
(dashed lines) as a function of the number of fluctuators for different
values of $\alpha$. The three pairs of lines (from top to bottom) 
refer to $\alpha=1.0$ (black), $\alpha=1.5$ (red) and $\alpha=2$ (blue).}\label{fig4}
\end{figure}

We now consider the more realistic case of a larger number of
fluctuators.  From Eq. \eqref{DNF}, and remembering that $0 \le {D} \le
1$, one sees immediately that the overall effect of having a large number of
fluctuators is to decrease the value of the optimal trace distance. As a
consequence, oscillations in the trace distance are damped and may
disappear, depending also on the value of $\alpha$, leading to a
monotonic decay. This behavior is illustrated in Fig. \ref{fig4}, where
$\nblp$ and $\nbcm$ are plotted as a function of the numbers of fluctuators 
and for three different exemplary values of $\alpha$.
The figure clearly shows that for smaller values of $\alpha$, i.e. 
in the $\alpha < 1$ regime, a small number of fluctuators is sufficient to completely
wash out memory effects. For increasingly larger values of $\alpha$,
non-Markovianity persists also for $N_f \approx 100$. Generally,
increasing the number of fluctuators brings the system towards a Markovian dynamics.
Summarizing, non-Markovianity is typical of environments with a small
number of fluctuators and a noise spectrum dominated by low-frequencies.  
\par
\begin{figure}[h!] 
\centering
\includegraphics*[width=0.9\columnwidth]{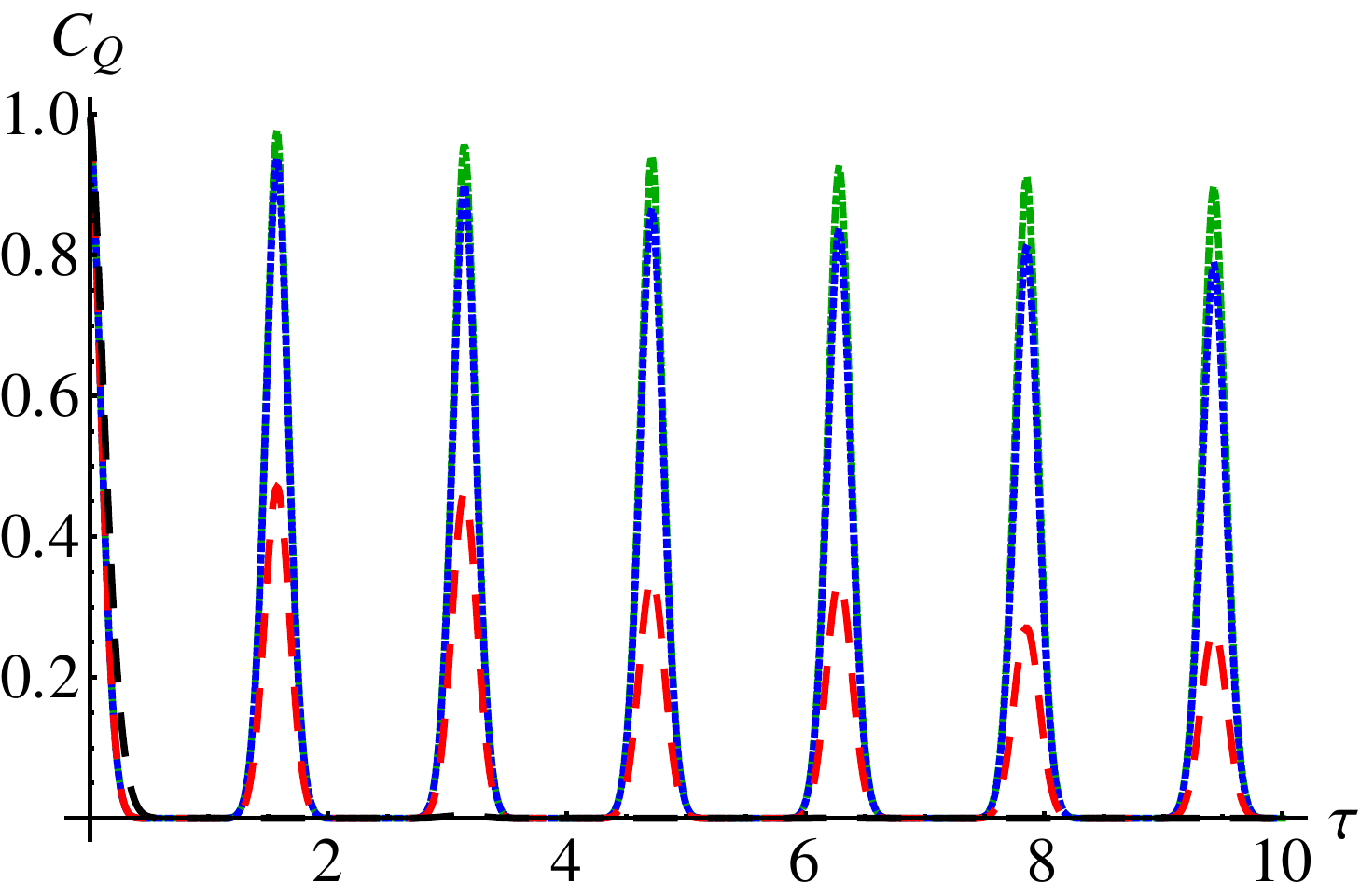}
\caption{(Color online) Non-Markovianity of colored channels. The plot
shows the quantum capacity for a qubit subject to
$1/f^{\alpha}$ noise generated by 10 random 
fluctuators for different values of $\alpha$: $\alpha=1$ (solid black line),
$\alpha=1.5$ (dashed red line), $\alpha=2$ (dotted blue line) and 
$\alpha=2.5$ (dotdashed green line).}\label{fig5}
\end{figure}

Let us now analyze the behavior of the quantum channel capacity.  In Fig.
\ref{fig5} we consider as an example the case of ten random fluctuators
and different values of $\alpha$. As expected, $C_Q(\tau,\alpha,N_f)$ is
an increasing function of $\alpha$. It has revivals at times multiples of
$\pi/2$.  As the number of fluctuators is increased the peaks become
narrower and smaller, and the $C_Q(\tau,\alpha,N_f)$ is zero almost
everywhere.  The quantum capacity is thus very sensitive to the channel
length or, equivalently, to the time during which the qubit is subjected
to noise. In more detail: only certain lengths of the channel,
corresponding to non-zero values of $C_Q(\tau,\alpha,N_f)$, allow for
reliable transmission of quantum information. This characteristic
lengths depend on the specific parameters of the noise. Besides, the range of
non-zero values of $C_Q(\tau,\alpha,N_f)$, decreases as $N_f$
increases, making robust quantum communication a more challenging 
task. {As mentioned above, these conclusions hold upon assuming that
the channel is reset after each use, i.e. focusing on memory 
effects during the propagation and neglecting memory effects among
subsequent uses of the channel.}
\subsection{Two-qubits non-Markovianity}
We conclude this section by addressing the non-Markovianity of RTN and
colored environments acting independently on two dephasing qubits.  The
dynamics is governed by the Hamiltonian \begin{equation} H(t)=
H_1(t)\otimes\mathbb{I}_2+\mathbb{I}_1\otimes H_2(t) \end{equation}
where $H_{1(2)}(t)$ is the single qubit Hamiltonian in Eq.
\eqref{hamiltonian} and $\mathbb{I}_{1(2)}$ is the identity operator in
the Hilbert space of the first (second) qubit.  If we focus on the BLP
measure, numerical maximization should be performed to find the
maximizing initial pair of states. In both the case of RTN and colored
noise, the maximizing pair corresponds to the two orthogonal factorized
states $|\text{++}\rangle$ and $|--\rangle$. We have numerically confirmed
that the optimal trace distance in this case also takes the form:
${D}=|\Gamma(t)|$. The BCM non-Markovianity measure is
straightforward to calculate as the quantum capacity, it is additive for
degradable channels as the one here considered, namely
$C_{2q}(\Phi(t))=2C_Q(\Phi(t))$, where the subscript $2q$ stands for two
qubits. It follows that the non-Markovian dynamics of two qubits
subjected to independent RTN or coloured noise is simply related to the
single-qubit non-Markovianity, and therefore it is quantitatively the
same.
\section{Conclusions}
We have addressed open quantum systems made of one or two qubits 
interacting with a classical random field and evaluated 
the non-Markovianity of the corresponding noisy maps.
In particular, we focused on environments showing non-Gaussian 
fluctuations, as those described by random telegraph noise 
and colored noise with spectra of the the form $1/f^\alpha$.  
Upon analyzing the dynamics of
both the trace distance and the quantum or entanglement assisted
capacity we have shown that the behavior 
of non-Markovianity based on both measures is qualitatively similar.  
Besides, we have shown that environments
with a spectrum dominated by low-frequency contribution are
generally non-Markovian, and that non-Markovianity of colored
environments decreases when the number of fluctuators realizing the
environment increases. 
\par
Overall, our results confirm that non-Markovianity may represent a
resource for quantum information processing. In particular, our results
show that non-Markovian features are indeed connected to the revivals of
quantum correlations. In fact, if we compare our results with those in Ref. 
\cite{benedetti13}, we find that whenever the environment is
non-Markovian, revivals of quantum correlations are present, while they
decay monotonically for a Markovian environment. In other words,
our results confirm that non-Markovianity cannot be considered as a mere 
label to identify different kinds of dynamics. Rather, it may be 
exploited for a better control of quantum channels, and to better preserve 
quantum correlations for quantum communication protocols and quantum 
information processing.  \label{sec4}
\section*{Acknowledgments}
We acknowledge financial support from the MIUR project FIRB-LiCHIS-
RBFR10YQ3H, the Emil Aaltonen Finnish foundation (Non-Markovian quantum
information project) and the Scottish Universities Physics Alliance
(SUPA). This work was carried out during a visit of CB, supported by EU
through the LLP Erasmus placement program, and MGAP, supported by SUPA
visiting professors programme, to Heriot-Watt University. CB and MGAP
thank F. Caruso, A. D'Arrigo, P. Bordone, F. Buscemi for useful 
discussions and Heriot-Watt University for hospitality.


\end{document}